\documentstyle[preprint,aps,12pt]{revtex}
\begin{document}
\draft
\date{\today}
\title{Spinors in Weyl Geometry}
\author{A.  H.  Fariborz \footnote{e-mail: AMIR@PINEAPPLE.APMATHS.UWO.CA}
        and
        D.  G.  C.  McKeon \footnote{e-mail: TMLEAFS@APMATHS.UWO.CA\\
                                        Fax: 519-661-3523}
}
\address{ Department of Applied Mathematics,\\
          University of Western Ontario,\\
          London, Ontario N6A 5B7,\\
          CANADA.\\} 
\maketitle
\begin{abstract}
We consider the wave equation  for spinors in ${\cal D}$-dimensional Weyl 
geometry. By appropriately coupling the Weyl vector $\phi _{\mu}$ 
as well as the spin connection $\omega _{\mu a b } $ to the spinor field, 
conformal invariance can be maintained.  The one loop effective action 
generated by the coupling of the spinor field
 to an external gravitational field is computed in two dimensions.  It
is found to be identical to the effective action for the case of a scalar 
field propagating in two dimensions.   
\end{abstract}
\newpage  
\section{Introduction}
The effective action for gravity generated by the coupling of 
an external gravitational field to quantum matter fields is of 
crucial importance in string theory [1,2].   
The behaviour of this induced gravitational theory under a 
conformal transformation is of
 particular interest.  In refs. [3,4],  the induced 
gravitational action in the case of a scalar matter field propagating 
in a background Weyl geometry is examined.  
In this paper, we extend these considerations to the case in which 
the matter field is a spinor field.

In the next section we consider how to couple a spinor field to a background 
gravitational field in the case of Weyl geometry in ${\cal D}$-dimensions.
We then compute the first Seely-Gilkey/DeWitt-Schwinger coefficient 
$E_1(x,{\cal D})$,
needed to determine the one-loop effective action in two dimensions when using
the formalism of ref. [5].
\section{Spinors in Weyl Geometry}   
Weyl in 1918 [6,7] introduced the idea of a conformal transformation
\begin{equation}
g_{\mu \nu } (x) 
\rightarrow
\Omega^2(x) g_{\mu \nu} (x)
\label{gmn_CT}
\end{equation}
on the metric tensor $g_{\mu \nu }(x)$. (We use the conventions of 
\cite{Adler}, except
that we employ the metric with signature ( + + + ...+) in ${\cal D}$ 
dimensions. )
In addition, he employed a vector field $\phi_\mu (x)$ in order to preserve
 the conformal invariance of the theory; the transformation of $\phi 
_\mu$ that accompanies that of (\ref{gmn_CT}) is
\begin{equation}
\phi_\mu \rightarrow \phi_\mu + \Omega ^{-1} \partial_\mu \Omega .
\label{phi_CT}
\end{equation}
As is explained in [6,8], the role of $\phi _\mu$ is to ensure that the 
magnitude of a vector $\xi ^\alpha $ ($ l^2=g_{\alpha \beta} \xi^\alpha
\xi ^\beta$ ) transforms as
\begin{equation}
dl=(\phi_\alpha dx^\alpha )l
\end{equation}
under a conformal transformation.   This in turn implies that the symmetric 
connection $\Gamma^\alpha _{\beta \gamma}$ is given by
\begin{equation}
\Gamma^\alpha_{\beta \gamma}=
-{1\over 2} g^{\alpha \lambda} 
( g_{\lambda \beta , \gamma} + g_{\lambda \gamma , \beta} - 
g_{\beta \gamma , \lambda} )
+
\delta ^\alpha _\beta \phi _\gamma + 
\delta ^\alpha _\gamma \phi _\beta -
g_{\beta \gamma} \phi^\alpha . 
\label{Connection}
\end{equation} 
The curvature tensor $R^\mu _{~ \nu \alpha \beta }$ is defined by 
\begin{equation}
\xi ^\mu _{~;\alpha ; \beta} - \xi ^\mu _{~; \beta ;\alpha}=
R^\mu _{~\nu \alpha \beta} \xi ^\nu
\end{equation}
where
\begin{equation}
\xi ^\mu _{~; \nu} = \xi^\mu_{~, \nu} - \Gamma ^\mu _{\lambda \nu} \xi ^\lambda
\end{equation}
and hence
\begin{equation}
R^\mu _{~\nu \alpha \beta} = 
- \Gamma ^\mu _{\nu \alpha , \beta }
+ \Gamma ^\mu _{\nu \beta , \alpha }
+ \Gamma ^\mu _{\lambda \beta} \Gamma^\lambda _{\nu \alpha}
- \Gamma ^\mu _{\lambda \alpha } \Gamma^\lambda _{\nu \beta} .
\end{equation}  
We also find from these equations that
\begin{equation}
g^{\mu \nu}_{~ ~;\lambda} = 
-2 g^{\mu \nu } \phi_\lambda ,
\label{gmn_diff_a}
\end{equation}
so that
\begin{equation}
g_{\mu \nu ; \lambda} = + 2 g_{\mu \nu } \phi _\lambda .
\label{gmn_diff_b}
\end{equation}
Consequently, we find that
\begin{eqnarray}
\xi^\mu_{~;\mu} &=& 
{
1
\over 
{\sqrt g} 
}
( {\sqrt g} \xi ^\mu )_{,\mu} - {\cal D} \phi_\mu \xi ^\mu 
\nonumber \\
&=& 
g^{\mu \nu } ( \xi _{\mu ; \nu } ) - ({\cal D} - 2 ) \phi_\mu \xi ^\mu
\label{Xi}
\end{eqnarray}
and therefore
\begin{equation}
\xi _{ \alpha ; \beta ; \gamma } - \xi _{\alpha ; \gamma ; \beta} =
R_{\alpha \eta \beta \gamma } \xi ^\eta +
2 \xi _\alpha ( \phi _{\beta ; \gamma} - \phi _{\gamma ; \beta } ).
\end{equation}
Since $\psi _{;\beta ; \gamma} - \psi _{; \gamma ; \beta} = 0 $, 
we see that if $ \psi = g_{\tau \lambda } \xi ^\tau \xi ^\lambda $then
\begin{equation}
\xi ^\alpha \xi ^\eta R_{\alpha \eta \beta \gamma} +
\xi ^2 
(\phi _{\beta ; \gamma} - \phi _{\gamma ; \beta} ) =0 
\end{equation}
and hence $R_{ \alpha \eta \beta \gamma } $is not anti-symmetric in the indices 
$\alpha $ and $\eta $.   

We now consider the vierbein field $e_{\mu a } $ defined so that
\begin{equation}
g_{\mu \nu } = \eta ^{ a b } e_{\mu a } e_{\nu b }
\end{equation}  
and
\begin{equation}
\eta_{ a  b } = g^{\mu \nu } e_{\mu a } e_{\nu b } .
\end{equation}  
By (\ref{gmn_diff_a}) and (\ref{gmn_diff_b}),  we see that
\begin{equation}
e^{\mu m }_{~~; \lambda} =
- e^{\mu m } \phi_\lambda
\label{e_diff_1}
\end{equation}
and 
\begin{equation}
e_{\mu m; \lambda } = e_{\mu m} \phi_\lambda .
\end{equation}
These equations, when combined with the definition of the covariant
derivative of $e^{\mu m}$,
\begin{equation}
e^{\mu m}_{~~;\lambda} = 
e^{\mu m}_{~~,\lambda} - \Gamma^{\mu}_{\kappa \lambda} e^{\kappa m}
+ \omega _{\lambda ~ n}^{~m} e^{\mu n }
\label{e_diff_2}
\end{equation}
show that the spin connection is given by
\begin{eqnarray}
\omega_{\lambda m a}& = &
-{ 1 \over 2} \left( e^\kappa_{~m,\lambda}e_{\kappa a} - 
e^\kappa_{~a, \lambda }e_{\kappa m} \right)  
-{1 \over 2} e^\kappa_{~m} e_{\lambda a , \kappa } +
{1 \over 2} e^\kappa _{~a} e_{\lambda m , \kappa} 
\nonumber \\
& &
+
{1\over 2} g_{\pi \lambda} 
         \left( e^\kappa _{~m} e^\pi_{~a,\kappa} - 
            e^\kappa_{~a} e^\pi_{~m,\kappa} \right)
-\phi_a e_{\lambda m} + \phi_m e_{\lambda a} .
\end{eqnarray}

Under the conformal transformations of (\ref{gmn_CT}) and 
(\ref{phi_CT}), we find that \begin{equation}
e_{\mu a} \rightarrow \Omega e_{\mu a}
\label{e_CT}
\end{equation}
\begin{equation}
\Gamma^\alpha_{\beta \gamma} \rightarrow \Gamma^\alpha_{\beta \gamma}
\label{Connection_CT}
\end{equation}
\begin{equation}
\omega_{\mu m n} \rightarrow \omega_{\mu m n }
\end{equation}
and 
\begin{equation}
R \rightarrow \Omega ^{-2} R
\label{R_CT}
\end{equation}
where $R=g^{\mu \nu } R_{\mu \nu}$ and $R_{\mu \nu}=
R^\lambda_{~\mu \lambda \nu}$.

A conformally invariant coupling between the gravitational field 
defined by $\phi_\mu$ and $ g_{\mu \nu}$ and a spinor field $\psi$
is given by 
\begin{equation}
S= \int d^{\cal D}x \, e {\bar \psi} e^{c \mu} 
\left[
i \gamma_c \left( 
                  \partial_\mu +{1\over 2} \omega _{\mu a b}
                  \sigma^{a b} + { {{\cal D}-1} \over 2} \phi_\mu
           \right)  
\right]
\psi
\label{S}
\end{equation}
where $\gamma_c$ is a set of ${\cal D}$-dimensional Dirac matrices satisfying
$\{ \gamma_a, \gamma_b \} = 2 \eta_{a b}$, and
$\sigma_{a b} = (1 / 4) \left[ \gamma_a,  \gamma_b \right]$.
The field $\psi$ undergoes the conformal transformation 
\begin{equation}
\psi \rightarrow \Omega ^{- \left( 
                              {{{\cal D}-1} \over 2} 
                           \right)
                          }
\psi
\end{equation}
in conjunction with (\ref{gmn_CT}) and (\ref{phi_CT}).  A term of the 
form $e {\bar \psi} {\sqrt R} \psi $ in (\ref{S}) would also be conformally
invariant, but will not be considered due to its non-polynomial 
dependence on $R$.

At one-loop order, the effective action for gravity due to the 
propagation of the spinor field $\psi$ in (\ref{S}) is given by 
$det D \! \! \! \! /$ where
\begin{equation}
D \! \! \! \! / =
e^{c \mu } \gamma_c \left(
                           \partial_\mu + {1\over2} \omega_{\mu a b}
                           \sigma^{a b } + { {{\cal D}-1} \over 2} \phi_\mu 
                    \right).
\label{D_slash}
\end{equation}
Since $D \! \! \! \! /$ is linear in the derivatives, we will replace
$ det D \! \! \! \! / $ by $det ^{ 1/2} \left( D\! \! \! \! / ^{~2} 
                                        \right) $. Despite the 
fact that $D \! \! \! \! /$ is not Hermitian due to the term 
proportional to $\phi_\mu$ in (\ref{S}), the anti-Hermitian part of 
$D \! \! \! \! / ^{~2}$ does not contribute in two dimensions as is
demonstrated by the following calculation in which 
$ D \! \! \! \! / ^{~2}$ is simplified.

If $\gamma_\mu = e_{\mu c} \gamma^c $ and 
$ D_\mu = \partial_\mu + {1\over 2} \omega_{\mu a b } \sigma^{a b}
+ {{{\cal D}-1}\over 2} \phi_\mu$, then by (\ref{D_slash}),
\begin{equation}
D\! \! \! \! / ^{~2} = \gamma^\mu 
                              \left( 
                                    \left[ D_\mu, \gamma^\nu \right] 
                                    +\gamma^\nu D_\mu  
                              \right) D_\nu
\label{D_slash2_1}
\end{equation}
with
\begin{equation}
\left[ D_\mu, \gamma^\nu \right] =
e^\nu_{~a,\mu} \gamma^a + 
{1\over 2} \omega_{\mu a b} e^\nu _{~c} 
\left[ \sigma^{a b}, \gamma^c \right]
\label{Dg_commute}
\end{equation}
where in both four and two dimensions 
\begin{equation}
\left[ \sigma_{a b}, \gamma^c \right] =
\gamma_a \delta^c_b - \gamma_b \delta^c_a .
\end{equation}
Together, (\ref{e_diff_1}), (\ref{e_diff_2}) and (\ref{Dg_commute}) yield
\begin{equation}
\left[ D_\mu, \gamma^\nu \right] =
\left[ \Gamma^\nu_{\mu \kappa}e^\kappa_p - e^\nu_p\phi_\mu \right]     
\gamma^p
\end{equation}
so that
\begin{equation}
D\! \! \! \! / ^{~2} =
\left[ 
e^\mu _q \left( \delta ^{q p}+ 2 \sigma^{q p} \right) 
         \left( \Gamma^\nu_{\mu \kappa} e^\kappa_{~p} - 
                e^\nu_{~p} \phi_\mu \right)
+ \left(  g^{\mu \nu} + 2 \sigma^{\mu \nu} \right) D_\mu 
\right]
D_\nu .
\label{D_slash2_2}
\end{equation}
We also can use the relation
\[
e^\mu_{~q} e^\kappa_{~p} \delta^{q p} \Gamma^\nu _{\mu \kappa} =
g^{\mu \kappa} \Gamma^\nu_{\mu \kappa}
\]
which by (\ref{Connection}) becomes
\begin{equation}
= g^{\nu \sigma}_{~~,\sigma} + {1\over 2} g^{\nu \lambda} g^{\mu \sigma}
g_{\mu \sigma , \lambda} + (2-{\cal D}) \phi^\nu .
\end{equation}
Consequently as
\begin{equation}
{1\over 2} g^{\mu \sigma} g_{\mu \sigma, \lambda} = 
{1\over {\sqrt g} }
{ 
{\partial {\sqrt g} } 
\over
{\partial x^ \lambda }
}
\end{equation}
we have 
\begin{equation}
D \! \! \! \! / ^{~2} =
\left[
       {1 \over {\sqrt g} } \left( 
                                  {\sqrt g} g^{\nu \lambda} 
                            \right)_{, \lambda}
       + (1-{\cal D}) \phi^\nu 
       - 2 \sigma^{q p} e^\mu_{~q} e^\nu_{~p} \phi_\mu 
       + \left( 
               g^{\mu \nu} + 2 \sigma^{\mu \nu} 
         \right) D_\mu
\right]  
D_\nu .
\end{equation}
We now find that as 
\begin{eqnarray}
\sigma^{\mu \nu} D_\mu D_\nu &=&
{1\over 2} \sigma^{\mu \nu} \left[ D_\mu, D_\nu \right]
\nonumber \\
&=& {1\over 2} \sigma^{\mu \nu}
\left[
       { {{\cal D}-1} \over 2 } \left( 
                                      \phi_{\nu,\mu} - \phi_{\mu,\nu} 
                                \right)
       + {1\over 2} \sigma^{a b} 
       \left( \omega_{\nu a b, \mu} -  \omega_{\mu a b, \nu} \right)
       + {1\over 4} \omega_{\mu a b}\omega_{\nu c d}
        \left[ \sigma^{a b}, \sigma^{c d} \right] 
\right]
\end{eqnarray}
with 
\[
\left[ \sigma_{a b}, \sigma_{c d} \right] = -
\left[
      \eta_{a c} \sigma_{b d} - \eta_{a d} \sigma_{ b c} +
      \eta_{b d} \sigma_{a c} - \eta_{b c} \sigma_{a d} 
\right],
\]
( and, in four dimensions 
$
\{ \sigma_{a b}, \sigma_{c d} \}=
{1\over 2} 
\left[ 
       -\eta_{a c} \eta_{b d} + 
       \eta_{a d} \eta_{b c} + \epsilon_{a b c d} \gamma_5
\right]
$)
we have
\begin{equation}
\sigma^{\mu \nu} D_\mu D_\nu =
{1\over 2} \sigma^{\mu \nu} 
\left[
       \left( { {{\cal D}-1} \over 2} \right) F_{\mu \nu} 
       -{1\over 2} \sigma^{a b} R_{\mu \nu a b}
\right].
\label{sigma_D_D}
\end{equation}
This involves use of the relations
$
R_{\mu \nu a b}=-\omega_{\nu a b, \mu} + \omega_{\mu a b, \nu}
                -\omega_{\mu a m}\omega^{~m}_{\nu ~ b}
                +\omega_{\nu a m}\omega^{~m}_{\mu ~ b}
$
with
$
R_{\mu \nu a b}= e^\alpha_{~a} e^\beta_{~b} R_{\mu \nu \alpha \beta}
$
and 
$
F_{\mu \nu}=\phi_{\nu ,\mu} - \phi_{\mu ,\nu}$.

We also have 
\begin{eqnarray}
-2 \sigma^{q p} e^\mu_{~q} e^\nu_{~p} \phi_\mu D_\nu&=&2 A^\nu {\dot D}_\nu
\nonumber \\
&=&{1\over {\sqrt g}}
\left[
       \left( {\dot D}_\mu + A_\mu \right) {\sqrt g} g^{\mu \nu}
       \left( {\dot D}_\nu + A_\nu \right)
       - {\dot D}_\mu \left( {\sqrt g} g^{\mu \nu} \right) {\dot D}_\nu
\right.
\nonumber \\
&&\left.   
   -\left( {\dot D}_\mu {\sqrt g} g^{\mu \nu} A_\nu \right)
       - {\sqrt g} A_\mu A^\mu
\right]
\label{A_D_dot}
\end{eqnarray}
where
\begin{equation}
A^\mu=\sigma^{\mu \nu} \phi_\nu
\end{equation}
and
\begin{equation}
{\dot D}_\mu \equiv \partial_\mu +{1\over 2} \omega_{\mu a b} \sigma^{a b},
\end{equation}
( so that $\left( {\dot D}_\mu {\sqrt g} g^{\mu \nu} A_\nu \right) =
\left( {\sqrt g} g^{\mu \nu} A_\nu \right)_{, \mu} +{1\over 2}{\sqrt g}
g^{\mu \nu} \left[ \omega_{\mu a b} \sigma^{a b}, A_\nu \right]$ ) as well as 
\begin{eqnarray}
\left[
      {1\over {\sqrt g} } \left( g^{\mu \nu} {\sqrt g} \right)_{, \mu} +
      (1-{\cal D})\phi^\nu +g^{\mu \nu} D_\mu 
\right] D_\nu
&=&
{1\over {\sqrt g} }
{\dot D}_\mu \left( {\sqrt g} g^{\mu \nu} \right) {\dot D}_\nu
-\left( {{{\cal D}-1}\over 2} \right)^2 \phi_\nu \phi^\nu 
\nonumber \\
& &
+\left( {{{\cal D}-1}\over 2} \right) {1\over {\sqrt g} } 
         \left( {\sqrt g} g^{\mu \nu} \phi_\mu \right)_{,\nu}.
\end{eqnarray}
Together (\ref{D_slash2_2}),(\ref{sigma_D_D}) and (\ref{A_D_dot}) reduce 
(\ref{D_slash2_1}) to 
\begin{eqnarray}
D\! \! \! \! / ^{~2} &=&
{1\over {\sqrt g}}
\left[
      \left( {\dot D}_\mu + A_\mu  \right) {\sqrt g} g^{\mu \nu}
      \left( {\dot D}_\nu + A_\nu  \right)
      - \left({\dot D}_\mu {\sqrt g} g^{\mu \nu} A_\nu \right)
      -{\sqrt g} A_\mu A^\mu
       - \left( {{{\cal D}-1}\over 2} \right) ^2 {\sqrt g} \phi_\mu \phi^\mu
\right.
\nonumber \\
& &
\left.
       + {{{\cal D}-1}\over 2} \left( {\sqrt g} g^{\mu \nu} \phi_\mu 
\right)_{,\nu} \right] 
+ \sigma^{\mu \nu} 
\left[
       \left( {{{\cal D}-1}\over 2} \right) F_{\mu \nu} -
        {1\over 2} \sigma^{a b} R_{\mu \nu a b}
\right].
\label{D_slash2_3}
\end{eqnarray} 
We note that $ D \! \! \! \! /^{~2}$ is not Hermitian due to the presence 
of the terms $ - \left( {\dot D}_\mu {\sqrt g} g^{\mu \nu} A_\nu \right)
        + {{{\cal D}-1}\over 2} \sigma_{\mu \nu} F^{\mu \nu} $ in 
(\ref{D_slash2_3}).  However, 
as we shall show in the next section, these terms do not contribute to the effective action
in two dimensions.

\section{The Effective Action for Gravity}
In order to compute the effective action for gravity induced by 
the propagation of a spinor in two-dimensional Weyl geometry, 
we use the formalism of ref.\cite{Bukhbinder}.
In this approach, we consider the operator
\begin{equation}
\Delta \! \! \! \! / = 
e^{1/2} \left( i D \! \! \! \! / \right) e^{-1/2}
\end{equation}
where $D \! \! \! \! / $ is defined in (\ref{D_slash}).  If we make the 
transforms of eqs. (\ref{gmn_CT}), (\ref{phi_CT}), 
(\ref{e_CT} - \ref{R_CT}) so that 
$\Delta  \! \! \! \! / \rightarrow {\bar \Delta }\!  \! \! \! / $, then 
it is easy to see that in all dimensions
\begin{equation}
\Delta \! \! \! \! / = 
\Omega^{1/2} {\bar \Delta }\! \! \! \! / \Omega^{1/2}
\end{equation}
and thus the formalism used to determine effective action for 
spinors propagating in a Riemannian background in ref.\cite{Bukhbinder} 
can be used in the case of a Weyl 
background; the one difference being that the expansion coefficient $E_1(x)$
given in (\ref{E1}) must be used in conjunction with the explicit forms for 
$V_\mu $ and $X$ that occur in (\ref{D_slash2_3}), viz.
\begin{equation}
V_\mu = {1\over 2} \omega_{\mu a b} \sigma^{a b} + \sigma_{\mu \nu} \phi^\nu
\end{equation}
and
\begin{equation}
X={1\over {\sqrt g}}
\left[
      \left( {\dot D}_\mu {\sqrt g} g^{\mu \nu} A_\nu \right) +
      {\sqrt g} A_\mu A^\mu +{1\over 4}{\sqrt g} \phi_\mu \phi^\mu
      -{1\over 2} \left( {\sqrt g} g^{\mu \nu} \phi_\nu \right)_{, \mu}
\right]
-{1\over 2} \sigma^{\mu \nu} 
\left( 
       F_{\mu \nu} - \sigma^{a b} R_{\mu \nu a b} 
\right).
\label{X}
\end{equation}
Since in two dimensions $\sigma_{\mu \nu}={1\over {2 i}} \epsilon _{\mu \nu}
\gamma_5$, $tr \left( \sigma^{\mu \nu} \right) =0$ and 
$
tr \left( \sigma_{\mu \nu} \sigma_{\alpha  \beta} \right) = 
-{1\over 2} \left( 
                   \delta_{\mu \alpha} \delta_{\nu \beta} - 
                   \delta_{\mu \beta}  \delta_{\nu \alpha}
            \right)
$
we see that 
\begin{eqnarray}
tr\left( E_1(x) \right)&=& {1\over {4 \pi}} tr 
\left[
      2 \left( {1\over 2} \omega_{\mu a b} \sigma^{a b} +
               \sigma_{\mu \nu} \phi^\nu 
        \right) \phi^\mu
      - {1\over {\sqrt g}} 
      \left[ 
            \left( 
                  {\dot D}_\mu {\sqrt g} g^{\mu \nu} A_\nu
            \right)
            + {\sqrt g} A_\mu A^\mu
              + {1\over 4} {\sqrt g} \phi_\mu \phi^\mu 
\right.
\right.
\nonumber \\
& &
\left.
\left.
              - {1\over 2} \left(
                                 {\sqrt g} g^{\mu \nu} \phi_\nu 
                           \right)_{,\mu}
       \right]
       +{1\over 2}  \sigma^{\mu \nu} \left(
                                            F_{\mu \nu} - 
                                            \sigma^{a b} R_{\mu \nu a b}
                                     \right)
      -{1\over 3}g^{\alpha \beta}\phi_{\alpha ; \beta}
      -{1\over 6} R                  
\right]
\nonumber \\
& = &{1\over {24 \pi}}
\left[
        R + 2 g^{\alpha \beta} 
       \left( \phi_\alpha \right)_{; \beta}
\right].
\label{Tr_E1}
\end{eqnarray}
Since in ${\cal D}$ dimensions [6,8]
$R={\dot R} + ({\cal D}-1)({\cal D}-2) \phi_\alpha \phi^\alpha- 
2 ({\cal D}-1) {1\over {\sqrt g}} \left( {\sqrt g} \phi^\alpha 
\right)_{,\alpha}$
where ${\dot R}$ is the Ricci  scalar in the limit $\phi_\alpha \rightarrow 0$,
$tr \left(  E_1 \right) ={1\over {24 \pi}} {\dot R}$, which is the 
Riemannian space limit.
(Those contributions to $X$ in (\ref{X}) that come from the non-Hermitian 
parts of
 $H$ have vanishing trace and hence do not contribute to the effective action,
 as was noted at the end of the preceding section.)

The result of (\ref{Tr_E1}) is precisely the same as the result that 
one obtains for
 a model in which a scalar propagates in a two-dimensional space with Weyl 
geometry [3,4].  Consequently, the arguments of \cite{Bukhbinder} can be 
used in the same way that they were in the scalar case, and we obtain  
\begin{equation}
S_{eff}= {1\over 24\pi}\int d^2 x \, {\sqrt g}
\left[
      -\sigma \left(
                     R +  { 2\over {\sqrt g}}
                    \left( {\sqrt g} \phi^\lambda \right)_{,\lambda} 
             \right)
      + \sigma \Box \sigma 
\right]
\label{S_eff}
\end{equation}
(where $ \sigma = ln \left( \Omega ^{-1} \right) $ ).  The integrand in 
(\ref{S_eff})
reduces to what occurs in Reimannian space in two dimensions as was 
argued above.  
Hence in two dimensions, the effective action for gravity induced 
by the propagation of a spinor in a background with Weyl geometry is identical to what one obtains with Riemannian background.
\section{Discussion}
We have analyzed in some detail the formalism appropriate for a spinor 
field propagating in a space-time with Weyl geometry.  Despite the fact that
the action for the spinor field is not Hermitian, we have shown that the 
anti-Hermitian contribution does not contribute to the effective action for 
gravity in two dimensions that is induced by the propagation of the spinor 
field.   Indeed, the spinor contribution turns out to be precisely equal to 
that of the scalar field in two dimensions.

Currently we are considering expanding our considerations to include a version
 of supergravity that involves Weyl geometry in the bosonic sector.
\acknowledgments
We would like to thank the Natural Sience and Engineering Research Council
of Canada for financial support.  R. and D. Mackenzie posed the questions
that stimulated this investigation.
\appendix
\section{}
In this appendix we consider the matrix element
\begin{equation}
M_{xy}=\langle x | e^{-H t } | y \rangle
\label{Mxx}
\end{equation}
for the operator
\begin{equation}
H=- {1\over {\sqrt g}}    \left( \partial _\mu + V_\mu \right) 
    {\sqrt g} g^{\mu \nu}  \left( \partial _\nu + V_\nu \right) + X.
\label{H}
\end{equation}
In order to compute the coefficient $E_1(x)$ in the expansion \cite{DeWitt} 
\begin{equation}
M_{xx} = {1\over {t ^{D/2}}}
         \left[
               E_0(x) + E_1(x) t + E_2(x) t^2 + ...
         \right]
\label{Mxx_expand}
\end{equation}
we employ the approach of [10,11] in conjunction with a normal coordinate 
expansion, as was employed in the non-linear sigma model 
\cite{Honerkamp}.

We begin by expanding the metric tensor in normal coordinate about a point $x$
to the order that will eventually be required to obtain $E_1(x)$, 
\begin{equation}
g_{\mu \nu} \left( x + \pi (\xi) \right) = 
g_{\mu \nu} + g_{\mu \nu ; \lambda} \xi^\lambda +
{1\over 2} 
\left[ 
       g_{\mu \nu ; \lambda \sigma} + 
       {1\over 3} \left(
                         R^\kappa_{~\lambda \mu \sigma}g_{\kappa \nu} +
                         R^\kappa_{~\lambda \nu \sigma}g_{\kappa \mu}
                  \right)     
\right]
\xi^\lambda \xi^\sigma + ...
\end{equation}
which, by (\ref{gmn_diff_a}), becomes
\begin{equation}
=g_{\mu \nu} + 2 g_{\mu \nu} \phi_\lambda \xi^\lambda +
{1\over 2} \left[ 
                 4 g_{\mu \nu} \phi_\lambda \phi_\sigma + 
                 g_{\mu \nu} \left(
                                   \phi_{\lambda ;\sigma}+\phi_{\sigma ;\lambda}                             \right) 
                +{1\over 3} \left( 
                                  R^\kappa_{~\lambda \mu \sigma} g_{\kappa \nu} 
                                 +R^\kappa_{~\lambda \nu \sigma} g_{\kappa \mu}
                            \right)
           \right]
\xi^\lambda \xi ^\sigma + ...
\label{gmn_expand}
\end{equation}
Consequently, we find that 
\begin{eqnarray}
{\sqrt {g \left( x + \pi (\xi) \right)} } &=&
exp \left[
           {1\over 2} tr \left[ ln (g_{\mu \nu}) \right]
    \right]
\nonumber \\
&=&{\sqrt g} 
\left[
      1+{\cal D} \phi . \xi +{{\cal D}^2\over 2} \left(\phi .\xi \right)^2 +
      {{\cal D}\over 2} \phi_{\lambda;\sigma} \xi^\lambda \xi^\sigma +
      {1\over 6} R_{\lambda \sigma}\xi^\lambda \xi^\sigma +...
\right]
\end{eqnarray}
and so
\begin{equation}
{1\over {\sqrt {g\left( x+ \pi (\xi ) \right)}}}= {1\over {\sqrt g}}
\left[ 
      1 - {\cal D} \phi.\xi +{{{\cal D}^2}\over 2} \left( \phi .\xi\right)^2 -
     {{\cal D}\over 2} \phi_{\lambda;\sigma} \xi^\lambda \xi^\sigma 
     -{1\over 6} R_{\lambda \sigma}\xi^\lambda \xi^\sigma +...
\right].
\end{equation}
We also will have occasion to use the expansions
\begin{equation}
V_\mu \left( x+\pi (\xi) \right) = 
V_\mu + V_{\mu ;\lambda} \xi^\lambda +
{1\over {2 !}} \left(
                       V_{\mu ;\lambda;\sigma} +{1\over 3} 
                       R^\kappa_{~\lambda \mu \sigma}V_\kappa 
                \right) \xi ^\lambda \xi^\sigma + ...
\label{Vm_expand}
\end{equation}
and
\begin{equation}
X\left( x+\pi (\xi) \right) = X + X_{;\lambda} \xi ^\lambda +
{1\over {2!}} X_{;\lambda;\sigma}\xi^\lambda\xi^\sigma + ...  
\label{X_expand}
\end{equation}
For purpose of illustration, we now let $X=V_\mu =0$ in (\ref{H}), so that
\begin{eqnarray}
H&=&-{1\over {\sqrt g}} \partial_\mu {\sqrt g} g^{\mu \nu} \partial_\nu
\nonumber \\
&=&\left( 
       1+Y_\lambda\xi^\lambda+Y_{\lambda \sigma}\xi ^\lambda \xi^\sigma  
 \right)
p_\mu 
\left(
       \delta^{\mu \nu} +Z^{\mu \nu}_\lambda \xi^\lambda +
       Z^{\mu \nu}_{\lambda \sigma}\xi^\lambda\xi^\sigma 
\right)
p_\nu 
\label{H_expand}
\end{eqnarray}      
to the order which we need to compute $E_1(x)$. (Here we have defined $p$ 
to be $-i \partial $ and have assumed $g_{\mu \nu}(x)=\delta_{\mu \nu}$.)
In order to expand $M_{xx}$ in (\ref{Mxx}) in powers  of $t$ as in 
(\ref{Mxx_expand}), we follow the 
approach of [10,11].  This involves first using the Schwinger expansion 
\cite{Schwinger}
\begin{eqnarray}
e^{-(H_0+ H_1) t}&=&e^{-H_0 t} +
(-t)\int_0^1 du \, e^{-(1-u) H_0 t}H_1 e^{-u H_0 t}
\nonumber \\
& &+ (-t)^2 \int_0^1 du\, u \int_0^1 dv \,
e^{-(1-u)H_0 t} H_1 e^{-u (1-v) H_0 t} H_1 e^{ -u v H_0 t} + ...    
\end{eqnarray}
where $H_0$ is independent of $\xi^\lambda$, and then converting this into 
a power series in $t$ by insertion of complete sets of momentum 
states and then performing the appropriate momentum integrals.

Explicitly,  we find that to the order that contributes to $E_1$, 
\begin{eqnarray}
M_{xx} & \approx & 
\langle x|
          \left[ 
                 (-t) \int_0^1 du \, e^{-(1-u)p^2 t}
                 \left( 
                      Y_{\alpha \beta} \xi^\alpha \xi^\beta p^2 +
                      p_\mu Z^{\mu \nu}_{\alpha \beta}\xi^\alpha \xi^\beta p_\nu                     +Y_\alpha\xi^\alpha p_\mu Z^{\mu \nu}_\beta \xi^\beta p_\nu                 \right)
                 e^{-u p^2 t}
          \right.
\nonumber \\
& &
+(-t)^2 \int_0^1 du \, u \int_0^1 dv \, e^{-(1-u)p^2 t}
\left(
      Y_\alpha \xi^\alpha p^2 + p_\mu Z^{\mu \nu}_\alpha \xi^\alpha p_\nu
\right)
e^{-u (1-v) p^2 t}
\nonumber \\
& &
\left.
\times
      \left(
             Y_\beta \xi^\beta p^2 + p_\lambda Z^{\lambda \sigma}_\beta
             \xi^\beta p_\sigma
      \right)
      e^{-u v p^2 t} + ...
\right]
| x \rangle .
\label{Mxx_E1}
\end{eqnarray}
If we now insert complete sets of position and momentum states into 
(\ref{Mxx_E1}) we obtain
\begin{eqnarray}
M_{xx} & \approx & (-t)\int_0^1 du \int {{dp dq}\over {\left( 2\pi 
\right)^{2  {\cal D}}}} 
\int dz \,  e^{-i(p-q).z} e^{-\left[ (1-u)p^2 +u q^2 \right] t}
\left( 
      Y_{\alpha \beta}z^\alpha z^\beta q^2 + p_\mu Z^{\mu \nu}_{\alpha \beta}
      z^\alpha z^\beta q_\nu
\right)
\nonumber \\
& &
+(-t) \int_0^1 du \int {{dp dq dr}\over {(2 \pi )^{3 {\cal D}}}}\int dz_1  
dz_2 \, e^{-i\left[ (p-q).z_1 +(q-r).z_2 \right] } 
e^{-\left[ (1-u)p^2+u r^2\right] t}
\left( Y_\alpha z_1^\alpha q_\mu Z^{\mu \nu}_\beta z_2^\beta r_\nu \right)
\nonumber \\
& &
+(-t)^2\int_0^1 du \, u \int_0^1 dv \int {{dp dq dr}\over {(2\pi )^{3 
{\cal D}} } } \int dz_1 dz_2 \, 
e^{-i \left[ (p-q).z_1 +(q-r).z_2 \right]}
e^{-\left[ (1-u)p^2 +u (1-v)q^2 +u v r^2\right] t}
\nonumber \\
& &
\times
\left(
      Y_\alpha z_1^\alpha q^2 + p_\mu Z^{\mu \nu}_\alpha z_1^\alpha q_\nu 
\right)
\left(
      Y_\beta z_2^\beta r^2 +q_\lambda Z^{\lambda \sigma}_\beta z_2^\beta 
      r_\sigma 
\right).
\label{Mxx_int}
\end{eqnarray}
It is now a straightforward exercise to first integrate over $z_i$ in 
(\ref{Mxx_int}), and 
then after utilizing the resulting delta functions, evaluate the momentum 
integrals over $p$,$q$ and $r$ (all of which are Gaussian). We finally arrive
 the result 
\begin{eqnarray}
M_{xx} &\approx &{t\over {( 4\pi t)^{{\cal D}/2}} }
\left[
      Y_{\alpha \alpha} \left(
                               -{{\cal D}\over 6} + {2\over 3}
                        \right)
      +\left(
             -{1\over 6} Z^{\mu \mu}_{\nu \nu}-{1\over 3} Z^{\mu \nu}_{\mu \nu}       \right)  
\right.
\nonumber \\
& &
+ Y_\alpha Y_\alpha ( {\cal D}+2 )\left(
                                 {{{\cal D}-4}\over 48}
                           \right)
+ Y_\mu Z^{\mu \nu}_\nu \left(
                              -{{\cal D}\over 12}+{1\over 3}
                        \right)
+ Y_\mu Z^{\nu \nu}_\mu \left(
                              {{{\cal D}-4}\over 24}
                        \right)
\nonumber \\
& &
\left.   
      +\left(
             {1\over 24} Z_\alpha^{\mu \nu}Z_\alpha^{\mu \nu}     
            +{1\over 12} Z_\lambda^{\mu \nu}Z_\nu^{\mu \lambda}
            +{1\over 48} Z_\lambda^{\mu \mu} Z_\lambda^{\nu \nu}
            -{1\over 12} Z_\mu^{\mu \lambda} Z_\lambda^{\nu \nu}
       \right)  
\right].
\label{Mxx_final}
\end{eqnarray}
If in (\ref{H_expand}) we were to include $X$ and $V_\mu$ as in (\ref{H}), 
then by employing the
expansions of (\ref{Vm_expand}) and (\ref{X_expand}), we find that in 
order to determine $E_1(x)$, (\ref{Mxx_final}) must be supplemented by
\begin{equation}
M_{xx} \approx {t\over {(4 \pi t )^{D/2}}}
\left(
      -V_\nu Y^\nu -X
\right).
\end{equation}
It is now possible to determine $E_1$ for the particular case in which 
$Y_\alpha$, $Y_{\alpha \beta}$, $Z^{\mu \nu}_\alpha $ and $Z_{\alpha 
\beta}^{\mu \nu}$ are fixed by the
expansions (\ref{gmn_expand} - \ref{X_expand}). We find that these imply 
that 
\begin{eqnarray}
Y_\alpha &=& - {\cal D} \phi_\alpha
\\
Y_{\alpha \beta} &=& {{{\cal D}^2}\over 2} \phi_\alpha \phi_\beta -{{\cal D} 
\over 4} \left(
                \phi_{\alpha ; \beta} + \phi_{\beta ; \alpha}
         \right)
-{1\over 6} R_{\alpha \beta}
\\
Z_\alpha^{\mu \nu} &=& ({\cal D}-2) g^{\mu \nu} \phi_\alpha
\label{Z}
\end{eqnarray}
and
\begin{eqnarray}
Z_{\alpha \beta}^{\mu \nu} &=& 
\left[
      g^{\mu \nu} \left(
                        {{{\cal D}^2}\over 2}- 2{\cal D} +2
                  \right)
      \phi_\alpha \phi_\beta + g^{\mu \nu} \left(
                                                  {{\cal D} \over 4}- {1\over2}
                                           \right)
      \left(
             \phi_{\alpha ;\beta} + \phi_{\beta ; \alpha}
      \right)
\right.
\nonumber \\
& &
\left.
     + {1\over 6} g^{\mu \nu} R_{\alpha \beta} 
      -{1\over 6} \left(
                         R^{\mu ~\nu}_{~ \alpha ~\beta} + 
                         R^{\nu ~\mu}_{~ \alpha ~\beta}
                  \right)      
\right].
\end{eqnarray}
Together, (\ref{Mxx_final} - \ref{Z}) imply that when ${\cal D}=2$,
\begin{equation}    
M_{xx} \approx {t\over {\left( 4 \pi t \right) }}
\left(
       2 V_\mu \phi^\mu -X -{1\over 3} g^{\alpha  \beta} \phi_{\alpha ;\beta}
       - {1\over 6} R
\right).
\label{Mxx_2D}
\end{equation}
By (\ref{Mxx_2D}) we see that in two dimensions
\begin{equation}
4\pi E_1(x) = 2 V_\mu \phi^\mu -X -{1\over 3} 
g^{\alpha \beta} \phi_{\alpha ;\beta} - {1\over 6} R
\label{E1}
\end{equation}
when considering the operator $H $ in (\ref{H}).

\end{document}